# Effects of Covid-19 Pandemic on Chinese Commodity Futures Markets

Ahmet Goncu

Department of Finance

International Business School Suzhou

Xi'an Jiaotong-Liverpool University

China, 215123

Email: Ahmet.Goncu@xjtlu.edu.cn

**Abstract**

In this study, empirical moments and the cointegration for all the liquid commodity futures traded in the Chinese futures markets are analyzed for the periods before and after Covid-19, which is important for trading strategies such as pairs trading. The results show that the positive change in the average returns of the products such as soybean, corn, corn starch, and iron ore futures are significantly stronger than other products in the post Covid-19 era, whereas the volatility increased most for silver, petroleum asphalt and egg futures after the pandemic started. The number of cointegrated pairs are reduced after the pandemic indicating the differentiation in returns due to the structural changes caused in the demand and supply conditions across commodities.

## 1. INTRODUCTION

Chinese commodity futures market is the global leader in trading volume for a wide range of commodities such as agricultural and industrial commodities[1]. Significant changes that occur in these markets due to the Covid-19 are likely to determine the future trend of the global commodity markets. The effect of the Covid-19 on the stock markets are documented in the studies such as Ambros et al. (2020), Harjoto et al. (2020), Cao et al. (2020). However, the effect of the pandemic is not uniform across different commodities and deserve more attention given their diversification and inflation hedging benefits for investors. Trading strategies such as pairs trading is highly dependent on the existence of cointegration between commodity futures. Therefore, we consider the Chinese commodity futures market to understand the global trends given China's leading role in the production and the consumption of these products.

---

[1] See 2015 WFE/IOMA Derivatives Market Survey reported by World Federation of Exchanges (WFE) and IOMA, "the commodity options and futures traded in Shanghai and Dalian accounting for 50% of the volume traded in 2015 in terms of number of contracts" (published, April 2nd, 2015).

Bakas & Triantafyllou (2020) empirically investigated the impact of the pandemic on the volatility of the US stock market and various commodity indices and show that the volatility of the commodity markets to a pandemic shock is associated with less uncertainty in prices. We document significant increase in the volatility of products such as silver and corn in the sample after the start of the pandemic. Differentiation across different commodity futures are observed while these effects are smoothed in commodity indices.

Amar et al. (2021) shows that there is spillover effect and co-movement after the Covid-19 among the financial markets of different countries while Chinese stock market is shown to be weakly integrated into the world markets possibly due to the fact that Chinese currency is not freely convertible. In Gharib et al. (2021), the relationship between international crude oil and gold prices are assessed in terms of contagion effects. Sakurai & Kurosaki (2020) showed that the upside and downside correlations between the oil and US stock markets is increased after the pandemic.

This study analyzes the changes in the empirical moments and co-integration across all the liquid Chinese commodity futures in the pre and post Covid-19 samples. Overall, it is observed that the co-movement among different commodity futures returns are weaker due to different supply and demand shocks across commodities. Especially, agricultural products show a statistically significant change in their average returns after the pandemic. Significant volatility, skewness and kurtosis changes are documented for various commodities.

---

[1] See 2015 WFE/IOMA Derivatives Market Survey reported by World Federation of Exchanges (WFE) and IOMA, "the commodity options and futures traded in Shanghai and Dalian accounting for 50% of the volume traded in 2015 in terms of number of contracts" (published, April 2nd, 2015).

## 2. DATASET

The dataset consists of thirty-two commodity futures traded in the Chinese markets utilizing the most active contract price for each trading day from 4-Jan-2016 to 31-Dec-2020 with 1,218 trading days. Thirty-two commodities are selected due to their high trading volume and sufficiently long trading history. In Figure 1, the scaled prices of the agricultural and industrial commodity indexes are plotted together with the Shanghai composite index (SSE) and gold prices. It is observed that in the post Covid-19 period agricultural index has the best Sharpe ratio.

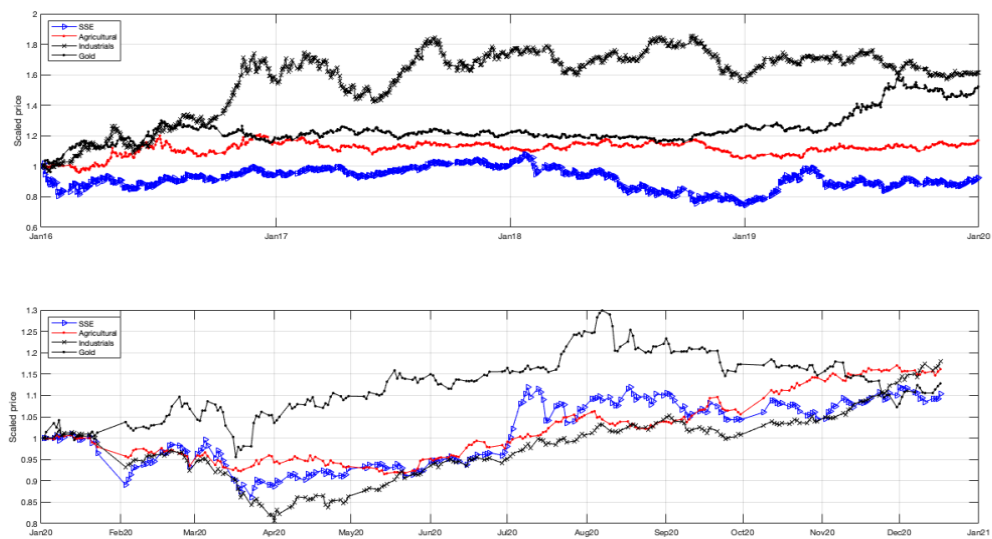

Figure 1 Plot of the equal weighted agricultural and industrial commodity indices, gold and stock market prices before and after the Covid-19 period. Initial prices are scaled to one.

In Table 1, the descriptive statistics of the commodity returns are given with two sample periods denoted as "before" and "after" Covid-19, respectively. For the agricultural commodities such as soybean, soybean meal, corn, corn starch, rapeseed dregs, egg, soybean oil, canola oil relatively there is a larger increase in average daily returns due to the pandemic, whereas sugar is an exception with lower average returns since most sugar demand is coming from the services and catering industries. A few of the products such as petroleum asphalt and PTA show lower average returns after the pandemic which are affected from the collapse of oil prices. Various other industrial and precious metals such as aluminium, copper, hot rolled coil, screw steel, nickel, zinc, iron ore, gold, and silver also show increased average returns in the post Covid-19 period.

---

[1] See 2015 WFE/IOMA Derivatives Market Survey reported by World Federation of Exchanges (WFE) and IOMA, "the commodity options and futures traded in Shanghai and Dalian accounting for 50% of the volume traded in 2015 in terms of number of contracts" (published, April 2nd, 2015).

**Table 1 Descriptive statistics of the commodity futures returns: before and after the Covid-19.**

|  | Mean | | Stdev. | | Skewness | | Kurtosis | |
| --- | --- | --- | --- | --- | --- | --- | --- | --- |
|  | Before | After | Before | After | Before | After | Before | After |
| soybean | 0.0001 | 0.0017 | 0.0109 | 0.0151 | 1.50 | -0.56 | 16.45 | 6.13 |
| silver | 0.0003 | 0.0011 | 0.0113 | 0.0267 | -0.27 | -0.30 | 9.02 | 4.61 |
| aluminium | 0.0003 | 0.0004 | 0.0099 | 0.0119 | 0.06 | -1.04 | 5.79 | 6.19 |
| gold | 0.0004 | 0.0005 | 0.0073 | 0.0118 | 0.40 | -0.34 | 8.07 | 6.47 |
| p. asphalt | 0.0006 | -0.001 | 0.019 | 0.027 | 0.59 | 0.42 | 11.39 | 8.45 |
| corn | 0.0000 | 0.0015 | 0.0149 | 0.0085 | -2.61 | 0.88 | 71.61 | 6.55 |
| cotton | 0.0002 | 0.0003 | 0.0134 | 0.0146 | -0.21 | -0.32 | 7.25 | 5.99 |
| corn starch | 0.0001 | 0.0012 | 0.0103 | 0.0096 | 0.38 | 1.11 | 15.22 | 8.16 |
| copper | 0.0003 | 0.0007 | 0.0101 | 0.0132 | 0.38 | -1.28 | 7.88 | 9.52 |
| glass | 0.0006 | 0.0009 | 0.0136 | 0.0151 | 0.07 | -0.59 | 6.00 | 5.55 |
| h. rolled coil | 0.0006 | 0.001 | 0.0188 | 0.0125 | -0.41 | -1.41 | 5.62 | 10.65 |
| iron ore | 0.0007 | 0.0017 | 0.0252 | 0.0244 | -0.96 | -0.92 | 9.21 | 8.50 |

---

[1] See 2015 WFE/IOMA Derivatives Market Survey reported by World Federation of Exchanges (WFE) and IOMA, "the commodity options and futures traded in Shanghai and Dalian accounting for 50% of the volume traded in 2015 in terms of number of contracts" (published, April 2nd, 2015).

| | | | | | | | | |
|---|---|---|---|---|---|---|---|---|
| coke | 0.0011 | 0.0017 | 0.0241 | 0.017 | -0.94 | -0.00 | 9.48 | 5.20 |
| egg | 0.0001 | 0.0007 | 0.0203 | 0.0348 | 4.36 | 2.46 | 76.24 | 26.07 |
| coal | 0.0008 | 0.0013 | 0.023 | 0.0209 | -0.83 | -1.65 | 9.37 | 22.89 |
| polythene | -0.0001 | 0.0002 | 0.0122 | 0.015 | -0.41 | -0.11 | 8.46 | 5.73 |
| soybean meal | 0.0002 | 0.0009 | 0.0122 | 0.0113 | -0.09 | 0.81 | 6.08 | 5.90 |
| m. alcohol | 0.0003 | 0.0004 | 0.0165 | 0.0192 | -0.10 | 0.27 | 4.52 | 5.37 |
| nickel | 0.0005 | 0.0004 | 0.0156 | 0.016 | -0.01 | -0.32 | 4.76 | 4.34 |
| canola oil | 0.0003 | 0.0009 | 0.0098 | 0.0132 | 0.11 | -0.74 | 6.46 | 6.14 |
| palm oil | 0.0003 | 0.0004 | 0.012 | 0.0187 | 0.17 | -0.35 | 5.43 | 3.54 |
| polypropylene | 0.0003 | 0.0003 | 0.014 | 0.0155 | -0.79 | -0.23 | 10.19 | 6.12 |
| screw steel | 0.0007 | 0.0009 | 0.0197 | 0.0118 | -0.36 | -1.15 | 5.58 | 11.21 |
| rapeseed dregs | 0.0002 | 0.0009 | 0.014 | 0.0147 | -0.17 | 0.51 | 5.32 | 5.27 |
| rubber | 0.0002 | 0.0003 | 0.0215 | 0.0278 | 1.62 | 0.60 | 19.14 | 20.21 |
| tin | 0.0004 | 0.0005 | 0.0107 | 0.0152 | 0.05 | -0.72 | 5.72 | 9.39 |
| sugar | 0.0000 | -0.0003 | 0.0087 | 0.0100 | 0.53 | 0.03 | 7.92 | 4.12 |
| pta | -0.0001 | 0.0011 | 0.013 | 0.0162 | -0.08 | -0.12 | 6.25 | 6.59 |
| p. chloride | 0.0003 | 0.0004 | 0.0127 | 0.0136 | -0.10 | -0.35 | 4.70 | 5.41 |
| soybean oil | 0.0002 | 0.0006 | 0.0095 | 0.0149 | 0.25 | -0.48 | 5.33 | 4.58 |
| steam coal | 0.0006 | 0.0008 | 0.016 | 0.0149 | -1.47 | -2.72 | 21.45 | 24.76 |
| zinc | 0.0003 | 0.0006 | 0.0136 | 0.0133 | -0.19 | -0.29 | 5.50 | 4.84 |

[1] See 2015 WFE/IOMA Derivatives Market Survey reported by World Federation of Exchanges (WFE) and IOMA, "the commodity options and futures traded in Shanghai and Dalian accounting for 50% of the volume traded in 2015 in terms of number of contracts" (published, April 2nd, 2015).

# 3. EMPIRICAL ANALYSIS

First, we compare the empirical distribution of the moments and the cointegration between all the commodity futures in our dataset for the pre and post Covid-19 periods. For the mean, standard deviation, skewness and kurtosis we calculate the difference between the pre and post Covid-19 samples. The standard deviation of these differences across thirty-two products are calculated to obtain the 90% confidence intervals for the changes in four sample moments and identify the products that have relatively larger changes in the sample moments due to the Covid-19.

The differences in the first four sample moments together with the confidence bands are presented in Figure 2. In the first subplot, soybean, corn, corn starch, and iron ore show large changes in their average daily log-returns due to the pandemic. These commodity futures showed very strong positive trend, whereas petroleum asphalt and PTA showed bearish pressure in their returns after the pandemic started. The figure clearly shows the upward shift in the average returns of almost all the commodities in the Chinese markets. Silver, petroleum asphalt, and egg futures show extreme increase in their volatility relative to other products. Volatility of corn, hot rolled coil, coke, and screw steel went down significantly more than the other products while most of the commodities showed higher volatility after the pandemic. The skewness of corn is significantly higher due to the supply shock, whereas the skewness for most of the other products decreased after the Covid-19. Finally, the sample kurtosis of corn and egg are significantly decreased after the pandemic, whereas it is increased for coal.

---

[1] See 2015 WFE/IOMA Derivatives Market Survey reported by World Federation of Exchanges (WFE) and IOMA, "the commodity options and futures traded in Shanghai and Dalian accounting for 50% of the volume traded in 2015 in terms of number of contracts" (published, April 2nd, 2015).

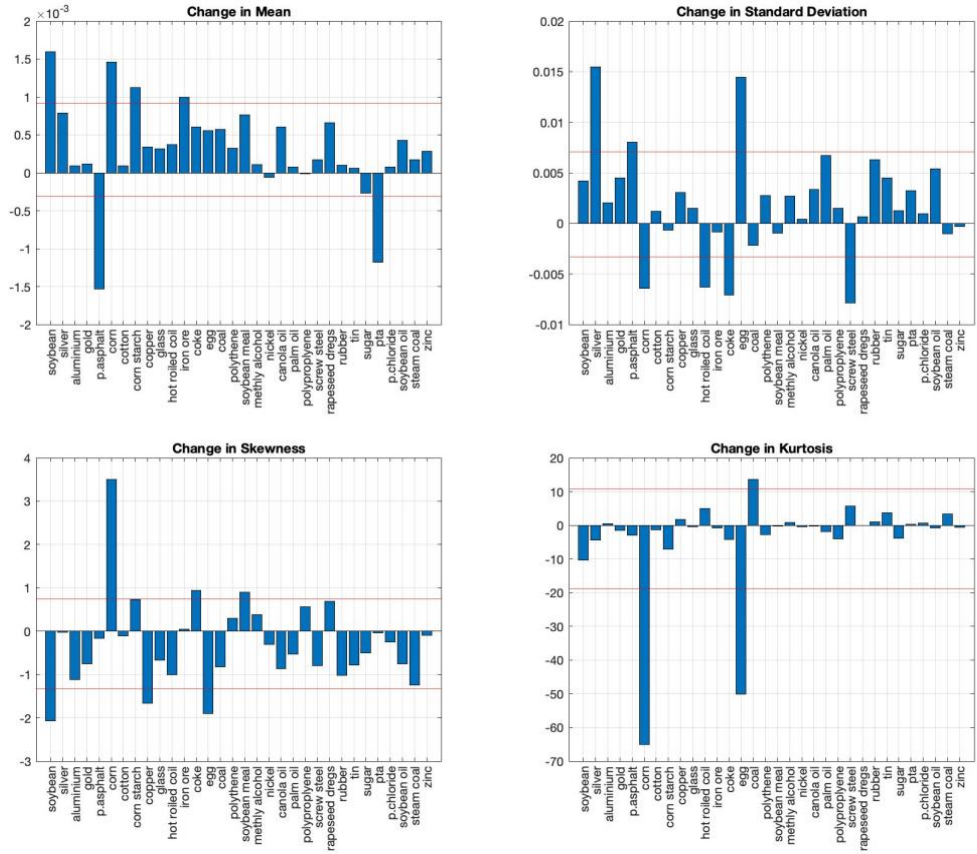

Figure 2 Changes in the empirical moments of commodity futures in the China.

Second, we analyse the cointegration between all the pairs in the dataset. Cointegration is used to find possible co-movements between time series processes in the long term, where the price series of two assets follow an equilibrium relationship. Engle-Granger cointegration test (see Engle & Granger, 1987) is employed with the null hypothesis (H$_0$) stating that there is no cointegration between the time series $P_t^B$ and $P_t^A$. The cointegration between the price series of A and B is given as:

$$\log(P_t^B) = a + b \log(P_t^A) + e_t, \qquad (1)$$

where $e_t$ is the i.i.d. Gaussian noise. The cointegration analysis is used to find the statistically significant changes in cointegration before and after the pandemic.

---

[1] See 2015 WFE/IOMA Derivatives Market Survey reported by World Federation of Exchanges (WFE) and IOMA, "the commodity options and futures traded in Shanghai and Dalian accounting for 50% of the volume traded in 2015 in terms of number of contracts" (published, April 2nd, 2015).

In Figure 3, cointegration results are given for each pair of commodities with the vertically and horizontally displayed names representing the series B and A in Equation (1), respectively. For each entry, the cointegration test results are given for the before and after the Covid-19 samples. The numbers "1" and "2" refer to the statistically significant cointegration at the 90% and 95% confidence levels, respectively. The value "0" means that we fail to reject the null hypothesis of "no cointegration". Yellow color indicates that the significant cointegration that existed before the pandemic is no longer significant, whereas the green color indicates the existence of significant cointegration that emerged after the pandemic. In Figure 3, there are 176 yellow 82 green values indicating that for most of the pairs the cointegration is lost in the post pandemic period and there are fewer number of cointegrated pairs. For example, gold, petroleum asphalt, PTA, sugar, and methyl alcohol are no longer cointegrated with any other commodity offering extra diversification benefits.

---

[1] See 2015 WFE/IOMA Derivatives Market Survey reported by World Federation of Exchanges (WFE) and IOMA, "the commodity options and futures traded in Shanghai and Dalian accounting for 50% of the volume traded in 2015 in terms of number of contracts" (published, April 2nd, 2015).

[Figure 3: Co-integration matrix table for 32 commodity futures, pre- and post-Covid-19 periods]

Figure 3 Co-integration analysis of thirty-two most liquid commodity futures products for the pre- and post- Covid-19 periods. The numbers "1" and "2" indicate that there is significant co-integration in the given period at the 10% and 5% confidence levels, respectively. For each product the pre- and post- Covid-19 period results are given.

## 4. CONCLUSION

All the liquid Chinese commodity futures are empirically analyzed in the pre- and post- Covid-19 periods. Various agricultural commodities and gold is less cointegrated with other products and performed well for the buy and hold strategies. Weaker cointegration between pairs of commodities indicate the investors should consider the commodity markets to hedge against inflation and diversify their investments. Especially, agricultural commodities are more important after Covid-19 as an investment for inflation hedging purposes. Electrification and green energy related commodities such as nickel, copper, silver, and zinc are showing improved average return performance. Therefore, investors should consider the new conditions of the commodity markets caused by the pandemic and adjust their asset allocations accordingly.

---

[1] See 2015 WFE/IOMA Derivatives Market Survey reported by World Federation of Exchanges (WFE) and IOMA, "the commodity options and futures traded in Shanghai and Dalian accounting for 50% of the volume traded in 2015 in terms of number of contracts" (published, April 2nd, 2015).

---

[1] See 2015 WFE/IOMA Derivatives Market Survey reported by World Federation of Exchanges (WFE) and IOMA, "the commodity options and futures traded in Shanghai and Dalian accounting for 50% of the volume traded in 2015 in terms of number of contracts" (published, April 2nd, 2015).

---

[1] See 2015 WFE/IOMA Derivatives Market Survey reported by World Federation of Exchanges (WFE) and IOMA, "the commodity options and futures traded in Shanghai and Dalian accounting for 50% of the volume traded in 2015 in terms of number of contracts" (published, April 2nd, 2015).